\documentclass[prl,aps,twocolumn,superscriptaddress,nofootinbib,floats,showpacs,10pt]{revtex4-1}

\usepackage{hyperref}
\usepackage{graphicx,epsfig}
\usepackage{amsfonts,amsmath,amssymb,amsthm,amscd}
\usepackage{mathptmx}
\usepackage{color}
\usepackage{dcolumn}
\usepackage{bm}
\usepackage{array}
\usepackage{float}

\newcommand{\av}[1]{\langle #1 \rangle }
\newcommand{\Prad}{P_\mathrm{rad}}

\begin{document}
\title{Depletion of intense fields}

\author{D. Seipt}
\affiliation{Helmholtz-Institut Jena, Fr{\"o}belstieg 3, 07743 Jena, Germany}
\affiliation{Lancaster University, Physics Department, Bailrigg, Lancaster LA1 4YW, UK}
\affiliation{Cockcroft Institute, Daresbury Laboratory, Keckwick Ln, Warrington WA4 4AD, UK}

\author{T. Heinzl}
\affiliation{School of Computing, Electronics and Mathematics, Plymouth University, Plymouth PL4 8AA, UK}

\author{M. Marklund}
\affiliation{Department of Physics, Chalmers University of Technology, SE-41296 Gothenburg, Sweden}

\author{S.~S. Bulanov}
\affiliation{Lawrence Berkeley National Laboratory, Berkeley, California 94720, USA}


\date{\today}

\begin{abstract}
The interaction of charged particles and photons with intense electromagnetic fields gives rise to multi-photon Compton and Breit-Wheeler processes. These are usually described in the framework of the external field approximation, where the electromagnetic field is assumed to have infinite energy. However, the multi-photon nature of these processes implies the absorption of a significant number of photons, which scales as the external field amplitude cubed. As a result, the interaction of a highly charged electron bunch with an intense laser pulse can lead to significant depletion of the laser pulse energy, thus rendering the external field approximation invalid. We provide relevant estimates for this depletion and find it to become important in the interaction between fields of amplitude $a_0 \sim 10^3$ and electron bunches with charges of the order of 10 nC.
\end{abstract}


\maketitle
 
The interaction of charged particles with ultra-intense electromagnetic (EM) pulses is the cornerstone of a newly emerging area of research, high intensity particle physics, located at the intersection of quantum electrodynamics (QED) and the theory of strong EM background fields. The latter significantly alter the physics of typical QED processes, leading to effects not encountered in perturbative quantum field theory \cite{reviews, ELI, NIMA, cascade, Reiss, Ritus}. Recently, there has been a surge of interest in these processes due to the planning and realization of new laser facilities, which will be able to deliver EM pulses of unprecedented intensities to test the predictions of high intensity particle physics \cite{ELI}. Moreover, the development of compact multi-GeV laser electron accelerators \cite{reviews,ELI,Esarey_RMP,LeemansPRL2014} adds another component necessary to carry out these studies.

Here, we will assume that the strong EM field is provided by an ultra-intense laser (pulse) with wave vector $k$, central frequency $\omega = 2\pi/\lambda$ in the optical regime and electric field magnitude $E$. The interactions of this strong field with photons and charged particles are parametrized in terms of the following parameters\footnote{We use natural units throughout: $\hbar = c = 1$.}: (i) the
(Lorentz and gauge invariant \cite{Heinzl:2008rh}) dimensionless amplitude of the EM vector potential, $a_0 = eE/\omega m$, (ii) the QED critical field, $E_S = m^2/e$ \cite{pair production}, (iii) the strong field invariants $\chi_e^2 = -e^2(F^{\mu\nu}p_\nu)^2/m^6$ and $\chi_\gamma^2 = -e^2 (F^{\mu\nu}k'_\nu)^2/m^6$ \cite{Ritus}. Here, $e$ and $m$ are electron charge and mass, $F_{\mu\nu}$ is the EM field tensor, while $p_{\nu}$ and $k'_{\nu}$ denote the 4-momenta of electron and photon probing the laser. The parameter $a_0$ is usually referred to as the classical nonlinearity parameter, since its physical meaning is the energy gain of an electron (in units $m$) traversing a reduced wavelength, $\lambdabar = 1/\omega$, of the field. For $a_0 > 1$ the electron/positron motion in such a field becomes relativistic. The parameter $E_S$ characterizes a distinct feature of QED, the ability to produce new particles {from vacuum}. This happens when an energy of $mc^2$ is delivered across an electron Compton wavelength, $\lambdabar_e = 1/m$, which is precisely achieved by $E_S$ \cite{pair production}. The parameters $\chi_e$ and $\chi_\gamma$ characterizes the interaction of charged particles and photons with the strong EM field. For example, $\chi_e$ is the EM field strength in the electron rest frame in units of $E_S$. Quantum effects become of crucial importance when $E \approx E_S$ or $\chi_{e,\gamma} \sim 1$. 

For large field amplitudes, $a_0 \gg 1$, the interaction of electrons/positrons and photons with strong EM fields involves the absorption of a large number of photons from the field.  Clearly, these correspond to an energy loss of the laser background field, which may or may not be negligible. Revisiting the results on multi-photon Compton and Breit-Wheeler processes \cite{Reiss, Ritus, C&B-W}, we find that there is indeed a parameter range, for which depletion of the laser becomes substantial. The processes in question have recently received a lot of interest \cite{C&B-W recent,Harvey:2009}, albeit with a focus on the \emph{final}  states (a frequency shifted photon or electron positron pairs). 

In this letter we want to change perspective and study in detail the dependence of nonlinear Compton scattering on the \emph{initial} multi-photon states, that is on the number of laser photons absorbed. This will allow us to establish a threshold for the validity of the external field approximation and discuss some immediate consequences. These findings should have a direct impact on the analysis of  QED backreaction on the classical EM field \cite{backreaction}. It should also be of great importance for the study of EM avalanches \cite{BellKirk, FedotovPRL2010, BulanovPRL2010}, since background depletion will significantly alter the energy partitioning of the processes. An avalanche is formed when Compton and Breit-Wheeler processes occur subsequently in an EM field of sufficiently high intensity, resulting in an exponential growth of the number of emitted particles.

The external field approximation is valid when the number of photons absorbed from the laser, $\Delta N_A$, is small compared to the total number $N_L \gg 1$ of photons in the pulse, which we take to be focussed to volume $V = \lambda^3$. A natural criterion for depletion is then provided by the equality $\Delta N_A = N_L$. The number of laser photons is proportional to intensity or field strength squared, $N_L \approx (2\pi/\alpha)(\lambdabar^2/\lambdabar_e^2)a_0^2 \approx 2 \times 10^{14}a_0^2$. Here $\alpha = e^2/4\pi \simeq 1/137$ denotes the fine structure constant. The number of absorbed photons is $\Delta N_A \simeq (\Delta E/\omega)N_T$, where $N_T$ is the number of electrons in the bunch and $\Delta E = P_\mathrm{rad} T$ is the energy loss upon radiating power $\Prad$ per laser period $T$. This power, a Lorentz invariant, can be estimated classically by making an analogy with synchrotron radiation \cite{Zeldovich:1975,Schwinger:1998}. In consequence, we will be able to estimate the number of photons absorbed from the field, the characteristic energy of an emitted photon and the angle of emission, implying a rather complete characterization of the processes.  To this end we go to a boosted frame, where the electron is on average at rest. If the laser is circularly polarized, the electron moves on a circle like in a synchrotron with 4-velocity $u = \gamma (1, \mathbf{\beta}_\perp, 0)$ where $\gamma^2 = (1 - \mathbf{\beta}_\perp^2)^{-1} = 1 + a_0^2$ characterizes the average rest frame (ARF). Using Larmor's formula, the radiated power becomes
$
 \Prad = - (2/3) \, \alpha \dot{u}^2 = (2/3) \, \alpha \, \omega^2 \, a_0^2(1+a_0^2) \; .
$
The boost to the ARF may be realized by choosing the initial electron momentum, $p = m \gamma_e  (1, 0,0, -\beta_e)$, such that its light-front component\footnote{If $\ell$ is an arbitrary four-vector its scalar product with the laser momentum can be written as $k \cdot \ell = \omega (\ell^0 - \ell_z) \equiv \omega \ell^-$, which defines the light-front component $\ell^-$ \cite{lightfront}.} equals $p^- =  m \gamma_e (1 + \beta_e) = m (1 + a_0^2)^{1/2} \equiv m_*$, with $m_*$ denoting the intensity-dependent effective mass \cite{EffMass}. At high energy, $\gamma_e \gg 1$, the radiation is emitted in the plane of electron motion, which in the ARF is perpendicular to the laser axis. In the lab frame this transforms into an emission angle
\begin{equation}\label{classical emission angle}
  \tan\theta = P_\perp/P_z = (2 a_0 m/ p^-)/[(m_*/p^-)^2 -1]  \; .
\end{equation}
determined by the ratio of longitudinal and transverse momenta in the ARF.
For $a_0 \ll 1$ the emission angle is $\sim 1/\gamma_e$, hence small, while for $a_0 \gg 1$ there is significant emission in the transverse direction. In the ARF, $\theta = \pi/2$. The number of absorbed photons per laser period $T = 2\pi/\omega$ is then
\begin{equation} \label{classical absorption}
  \Delta N_A = (4\pi/3) \, \alpha \, a_0^2(1+a_0^2) \, N_T.
\end{equation} 
So for $a_0 \gg 1$, the radiated power, hence the number of absorbed photons per laser cycle, increases like $a_0^4$. From synchrotron radiation it is known that the power radiated into the $s$-th harmonic asymptotically scales like $P_s \sim s^{1/3}$ \cite{Schwinger:1998}, so that the total power is $P = \sum_s^{s_0} P_s \sim s_0^{4/3} \sim a_0^4$. We thus obtain the important result that the typical number $s_0$ of laser photons, absorbed to yield emission of a single high-energy photon, scales like $s_0 \sim a_0^3$.

Turning back to the question of beam depletion, we equate $\Delta N_A {\sim} N_L$ to see that depletion requires
\begin{equation} \label{classical depletion}
  a_0^2 N_T \sim 6.5 \times 10^{15}.
\end{equation}
For an electron bunch containing a charge of 1 nC, a laser with $a_0\approx 10^3$ is needed. 
For such values of $a_0$ the energy $\omega'$ of the emitted photons is of the order of the electron energy gain per laser period, and the emission angle significantly deviates from $\sim 1/\gamma_e$. Thus, in this case, one expects not just significant radiation reaction with ensuing changes of the particle trajectories \cite{RR} but also strong recoil of the electron momentum. These features are best described in quantum theory to which we now turn. 

\begin{figure}[tb]
\includegraphics[width=0.99\columnwidth]{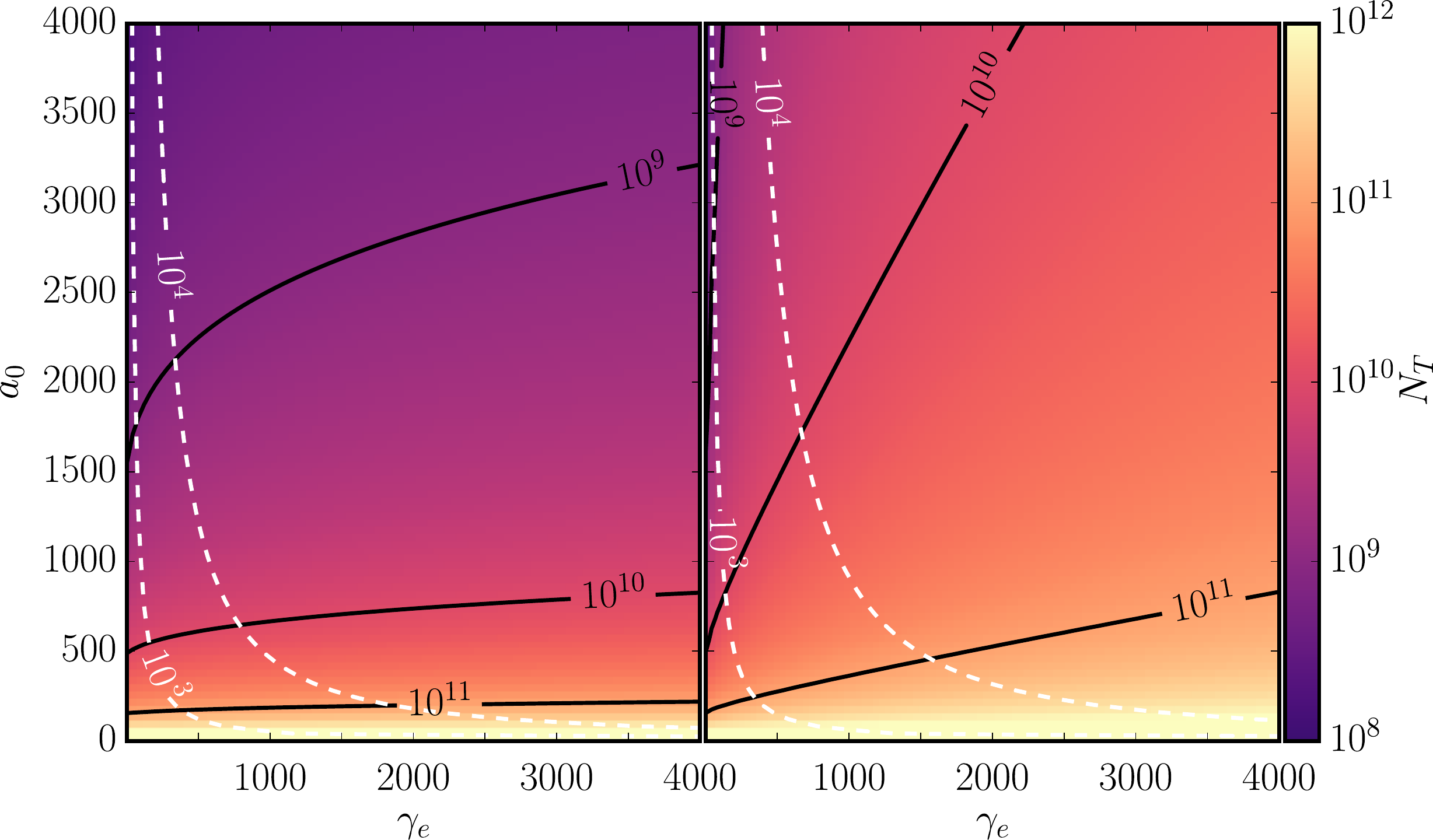}
\caption{
{The number $N_T$ of radiating electrons required to see significant depletion plotted as a function of $\gamma_e$ and $a_0$. {Black curves indicate the depletion thresholds $a_0 (\gamma_e)$ when $N_T=10^{9}$, $10^{10}$ and $10^{11}$. The dashed white curves represent $\omega'\Delta x = const \gg 1$, so that photon emission is incoherent across the relevant parameter space. Left: using \eqref{depletion}. Right: using \eqref{eq:dPe_ds} below.}}
}
\end{figure} 

First, we estimate the depletion threshold by taking into account the discrete nature of photon emission. While the average number of absorbed photons, $s_0$, still follows the classical scaling law $s_0 \sim a_0^3$ for $a_0 \gg 1$ \cite{Ritus}, the classical formula (\ref{classical absorption}) is replaced by
\begin{equation}\label{depletion}
  \Delta N_A \sim s_0 N_T (\lambda/L_C) \sim a_0^3 N_T (\lambda/L_C) \; .
\end{equation} 
Here, $L_C$ is the radiation length of the electron in a strong EM field \cite{Ritus} so that, on average, there is one photon emission per distance $L_C$. The classical behavior (\ref{classical absorption}) is recovered in the limit $\chi_e \ll 1$ where $L_C \sim \lambda /a_0$. In the deep quantum regime, $\chi_e \gg 1$, we employ the asymptotic expression 
$L_C = 0.43 \lambda \gamma_e^{1/3} a_0^{-2/3}$ \cite{Ritus} to obtain a quantum formula for the threshold of depletion,
\begin{equation} \label{depletion estimate}
    a_0^2 N_T/(a_0 \gamma_e)^{1/3} \sim   10^{14} \; , 
\end{equation}
which supersedes (\ref{classical depletion}). Again, for an electron beam of 1 nC and $\gamma_e \simeq 10^3$, a laser with intensity $a_0 \approx 10^3$ is required. Intensities of this magnitude should become reality in the near future \cite{ELI}. 
The critical value of $a_0$ depends weakly on the initial electron energy ($\sim \gamma_e^{1/5}$) as shown in Fig.~1, where $a_0(\gamma_e)$ is shown for different values of $N_T$. Thus, taking quantum effects into account increases the critical value of $a_0$ needed to deplete the laser pulse for a given value of initial electron momentum. 
We note that, for $\chi_e \gg 1$, the depletion of the electron beam energy is quite strong \cite{e-beam depletion}. This corresponds to the threshold for depletion of the laser beam going down from (\ref{depletion estimate}) to (\ref{classical depletion}), as illustrated in Fig.~1. 
Both (\ref{classical depletion}) and (\ref{depletion estimate}) show that, when a sufficiently charged electron bunch collides with an intense laser pulse, depletion of the laser pulse can become significant, with the originally strong EM field turning weak.  The required number of electrons is quite typical for an EM avalanche \cite{FedotovPRL2010}, where an intense laser produces a copious amount of high energy photons and subsequently electron-positron pairs.
{As the required electron densities are quite substantial, we briefly address the issue of coherence effects. {To this end we note that the inter-electronic distance $\Delta x \sim \lambda/\sqrt[3]{N_T}$ remains much larger than the emitted photon wavelength $\propto 1/\omega'$.} As a result, $\omega'\Delta x\gg1$ (see Fig.~1), a well established criterion for hard photon emission to be \emph{in}coherent~\cite{hartemann}.}

Second, we refine the depletion threshold estimate by calculating the quantum corrections to the average number 
of absorbed photons taking into account the probabilistic nature of photon emission.
{In line with the current understanding of high-intensity laser matter interactions in the quantum regime, we model photon emission as succession of incoherent one-photon events \cite{PIC,PIC-review,BulanovPRL2010,incoherent_model}.
There will be regimes where this assumption becomes challenged, for instance when extreme field strengths are reached such that $\alpha \chi^{2/3} \gtrsim 1$. In this case, higher-order diagrams such as self-energy corrections \cite{loops} and coherent multi-photon emission \cite{doublecompton} can no longer be neglected.
{A detailed investigation of higher-order effects is clearly beyond the scope of the present study, but we can at least state that $\alpha \chi^{2/3} \lesssim 0.1$ for our parameter range.}}
{It is thus sufficient to} introduce {one-}particle emission probabilities $d P^{\gamma,e}/ds$, which are differential in the number of photons $s$ absorbed from the laser field.  The average amount of energy $\av{\mathcal{E}}$ drawn from the laser field in a single photon emission or pair production is then $\av{\mathcal{E}} = \omega \av{s}$, with the average number of absorbed laser photons given by the
expectation value $\av{s}_{e,\gamma} = Z^{-1} \int \! ds \, s \, (dP^{e,\gamma}/ds)$ with with normalization integral $Z = \int ds \, dP^{e,\gamma} / ds$.

In a monochromatic plane wave laser field, taken to be circularly polarized for simplicity, the variable $s$ is discrete and describes the emission of higher harmonics due to absorption of $s$ laser photons. 
{Introducing the usual } {\emph{quasi} momentum $q = p + (m^2 a_0^2/ 2 k\cdot p) k$ (and analogously for $q'$, whence $q^2 = q'^2 = m_*^2$), kinematics become encoded in quasi momentum conservation, $q + sk = q' + k'$.}
The partial probabilities (per unit time), $P_s^{e}$, were calculated long ago 
\cite{Ritus, C&B-W} and give the total probability for Compton photon emission when summed over all harmonics: $ P^e=\sum_{s=1}^\infty P^e_s$. For large values of $a_0\gg 1 $,
the number $s$ of harmonics contributing grows like $s \sim a_0^3$, hence can be  
assumed quasi-continuous.
The sum may thus be replaced by an integral over $s$ with integrand {\cite{Ritus}}

\onecolumngrid
\begin{align}\label{eq:dPe_ds}
\frac{dP^e}{ds}  = \frac{4\alpha \omega s}{1+a_0^2}  \left( \frac{2}{s}\right)^{2/3}
\intop_0^1 \frac{dt}{(1 + s u_1 t)^2}
\left\{
 - \Phi^2( \eta ) + a_0^2 \left( \frac{2}{s} \right)^{2/3} 
 \left( 1 + \frac{s^2 u_1^2 t^2}{2 (1+su_1t)} \right)
 \left[
 \eta \Phi^2(\eta ) + \Phi'(\eta )^2
 \right]
\right\}\,.
\end{align}
\twocolumngrid

$\Phi$ and $\Phi'$ denote the Airy function and its derivative, their argument being $\eta(s,t) = \left( s/2 \right)^{2/3} \left[ 1 - 4 a_0^2 (m/m_*)^2 \, t (1-t)  \right]$, with $t = u/(s u_1)$, $u= (k'\cdot k)/(p'\cdot k)$ and $u_1 = 2(k\cdot p)/m_*^2$.
The dependence of $s dP^e/ds$ on $s$ is shown in {Fig.~2b}. The maximum corresponding to the most probable number of absorbed photons shifts towards lower values of $s$ with increase of initial electron energy. In {Fig.~2a} the dependence of $\av{s}_e$ on the parameter $a_0$ shows an increase of the number of absorbed photons with the EM field strength, but indicates a dependence different from the classical behavior, $s_0 \sim a_0^3$: the most probable number of absorbed photons also depends on the parameter $\chi_e$ as given by the fit $\av{s}_e=0.54a_0^3/(1+1.49\chi_e^{0.59})$. Using $\chi_e = 2\gamma_e a_0 \omega/m$, the threshold for depletion becomes
$
a_0^{1.08} \gamma_e^{-0.92}N_T\sim 6.8\times 10^{11}.
$  
For instance, when $a_0 = 10^3$, we find a value of $N_T \approx 10^{11-12}$, cf.\ Fig.~1 (right), which is larger than $N_T \approx 10^{10}$ predicted by the simple estimate (\ref{depletion estimate}), but still within reach of EM avalanches \cite{FedotovPRL2010}.

We note that for classical synchrotron emission it is straightforward to relate radiated to absorbed power, because of the continuity of emission. In the quantum case a typical interaction of an electron beam with an intense laser pulse proceeds via multiple emissions, each of them potentially resulting in a significant change of the electron momentum. To characterize such interactions one uses simulation codes with QED modules, which take into account multi-photon Compton and Breit-Wheeler processes. For these codes to be used for depletion calculations, each Compton process needs to be characterized by photon energy, angle of emission and the number of absorbed photons. Furthermore, in numerical (QED-PIC) simulations of multi-stage emission processes, which lead to the formation of avalanches/cascades, photon and electron emission angles strongly determine the probability of the subsequent pair production or photon emission process, respectively \cite{cascade,BulanovPRL2010,PIC,collapse,PIC-review,Jirka}. We hence proceed by calculating these quantities. 

The probabilities $dP^e/ds$ determine the number distribution of photons absorbed from the laser field in a single high-frequency photon emission. In what follows, we relate $dP^e/ds$ to the distribution $dP^e/d\chi_\gamma$ of the scattered photon longitudinal momentum ($\chi_\gamma  \sim k \cdot k^\prime$), which determines the intensity of the emitted high-energy photon radiation, via the chain rule:
\begin{align}\label{chain rule}
\frac{dP^e}{d\chi_\gamma} = \frac{ds(\chi_\gamma)}{d\chi_\gamma} \, \frac{dP^e}{ds} \; ,
\end{align}
where the functional relation $s = s(\chi_\gamma)$ is unknown. From the $t$-integral in \eqref{eq:dPe_ds} we see that the integrand is sharply peaked at $t=1/2$. Using energy momentum conservation we can solve $t = u/(su_1) = 1/2$ for $s$ with the result 
\begin{align}\label{s}
  s(\chi_\gamma) = \frac{a_0^3}{\chi_e} \frac{\chi_\gamma}{\chi_e-\chi_\gamma}  \, .
\end{align}
{This is valid for $a_0 \gg 1$ and reproduces the leading order of the related result (18) in \cite{Narozhnyi:1965}.}
A direct \emph{numerical} determination of $s(\chi_\gamma)$ from (\ref{chain rule}) shows excellent agreement with \eqref{s} for the most important range of $s\sim a_0^3$ (but deviates for $s\to 0$).
Thus, when a Compton photon with a given value of $\chi_\gamma$ is emitted, the number of laser photons drawn from the laser field can safely be estimated using \eqref{s} within the model of one-photon incoherent emission\footnote{A formula completely analogous to \eqref{s} holds for the Breit-Wheeler process, $\gamma + s\gamma_L \to e^+ e^-$, which becomes possible above a threshold in photon number, $s \geq s_0 = 2a_0(1+a_0^2)/\chi_\gamma$. Details will be discussed elsewhere.}.

\begin{figure}[!th]
\includegraphics[width=0.99\columnwidth]{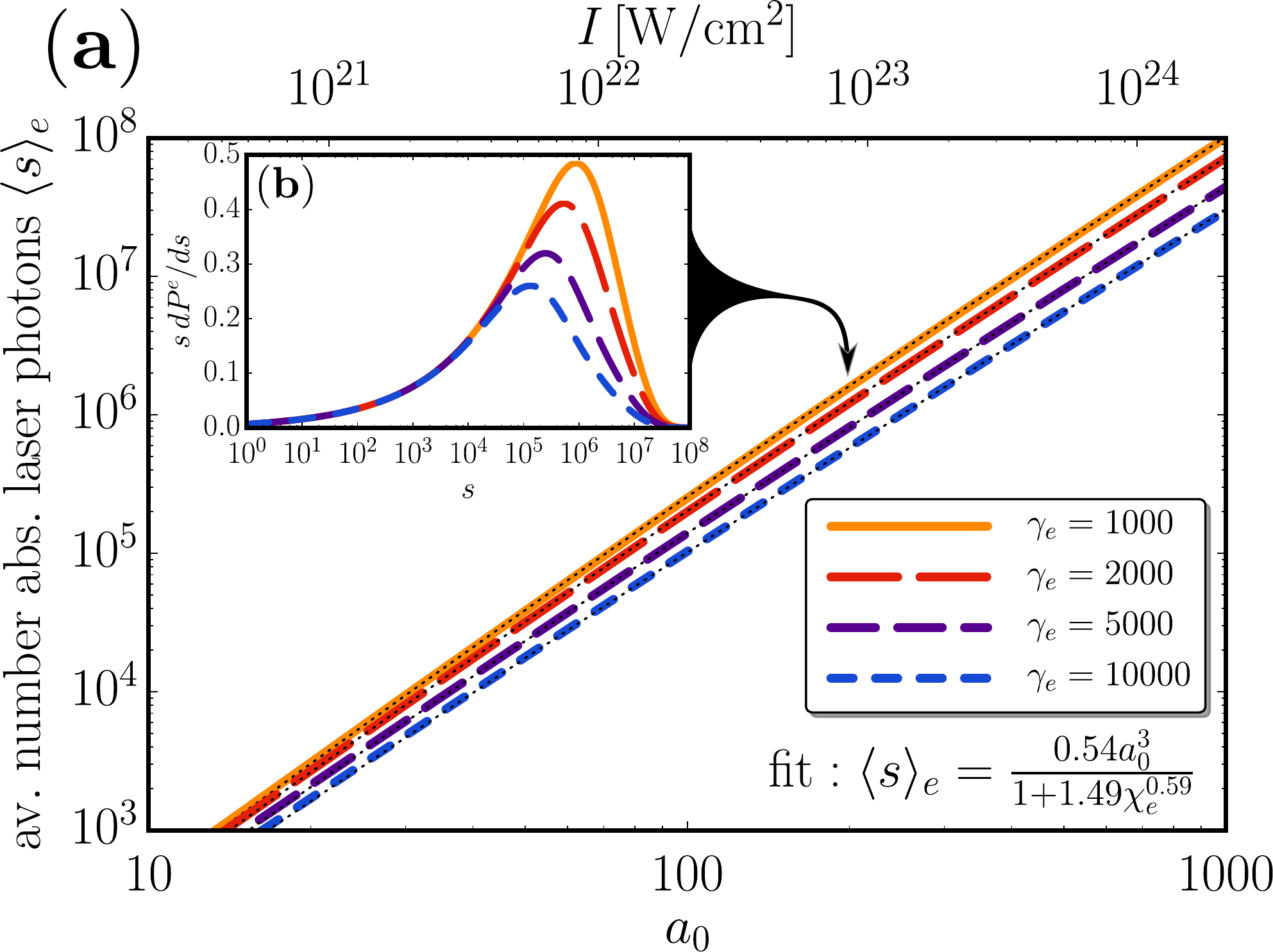}
\caption{
{The dependence of the average number $\av{s}_e$ of absorbed laser photons on the dimensionless amplitude $a_0$ of the EM field
for different values of $\gamma_e$. The corresponding distributions $s dP^e/ds$ for $a_0=200$ are shown in the inset (b).
Black dotted curves represent the numerical fit provided.}}
\end{figure} 

To further illustrate the power of the result \eqref{s}, we employ it to determine the most probable emission angle without referring to an angular probability distribution.  Let us write the scattered photon momentum as $k' = (\omega', \mathbf{k}'_\perp, k'_z)$ where $k'^2 = 0$. We can then find $k'_\perp$ from quasi-momentum conservation. Assuming a head-on collision of electrons and laser ($\mathbf{p}_\perp = 0$) the following answer is obtained:
\begin{equation} \label{k.perp.s}
  k'^2_\perp = 2s k \cdot k' - \left( \frac{k \cdot k'}{k \cdot p} \right)^2 \left(m_*^2 + 2s k \cdot p\right) \; .
\end{equation} 
This identity is manifestly invariant with respect to boosts collinear with the laser direction $k$. It defines an ellipse in the $(k^\prime_z, k_\perp^\prime)$ plane for given values of $\gamma_e$, $s$ and $a_0$, see Fig.~3.
Plugging (\ref{s}) into (\ref{k.perp.s}) yields the tangent of the most probable photon emission angle,
\begin{equation} \label{tan-theta}
  \tan \theta = k'_\perp/k'_z = \frac{4a_0 \gamma_e}{4 \gamma_e^2 - a_0^2} \; ,
\end{equation}
where $a_0, \gamma_e \gg 1$. 
This coincides with the classical emission angle (\ref{classical emission angle}) and is indeed consistent with the findings of \cite{Harvey:2009}: As long as $\gamma_e \gg a_0$, the photons are predominantly emitted in the forward direction, with $\theta\sim a_0/\gamma_e \ll 1$. However, as $a_0$ increases, significant photon emission takes place in the perpendicular direction. This can be understood classically, in particular in the ARF where  $a_0 \simeq 2\gamma_e$ ($p^- = m_*$), so that $\theta = \pi/2$ as required for circular (synchrotron) motion in the transverse plane as well as by \eqref{tan-theta}. Equivalently, this follows  from the classical equation of motion by calculating $\tan \theta = (\pi_\perp/\pi_z)_\mathrm{rms}$, the ratio of the rms values of the classical electron momentum components in the laser field $A$, $\pi^\mu = p^\mu - e A^\mu + (e p\cdot A - e^2 A^2/2) \, k^\mu/k\cdot p$.

\begin{figure}[tb]
\includegraphics[width=0.99\columnwidth]{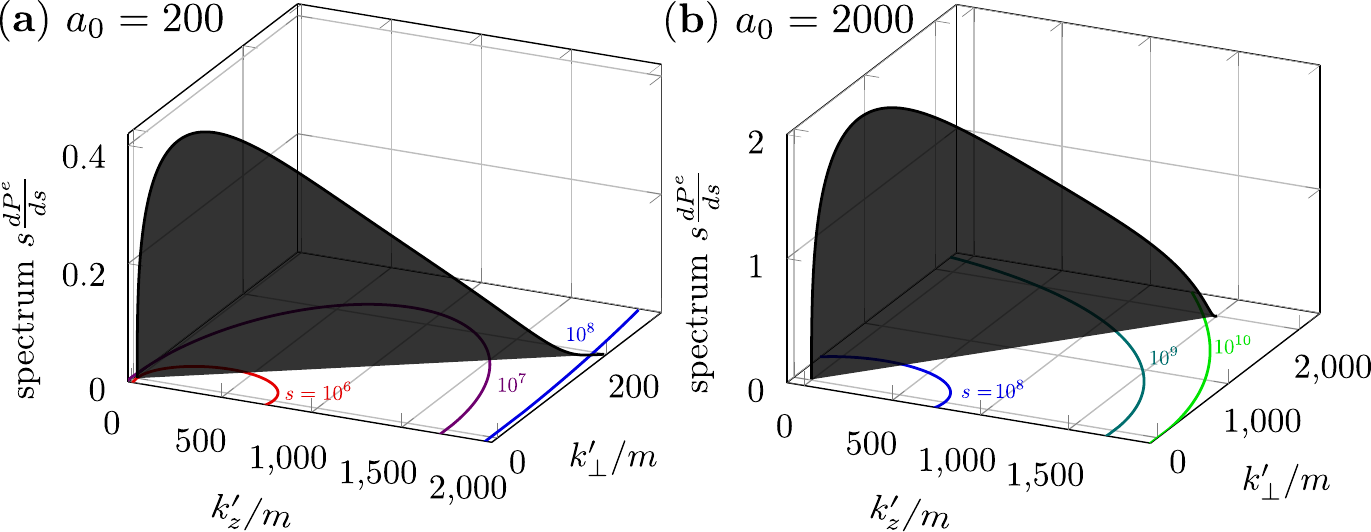}
\caption{
The distribution $s d P^e / d s $ as a function of  $k'_\perp$ and $k'_z$ 
for $\gamma_e= 2000$, and for $a_0=200$ (left) and $2000$ (right). The distributions are
supported on a line $k'_\perp/k'_z = 1/10$ and $4/3$ in the left and right panels, respectively, cf.\ (\ref{tan-theta}).
Full curves on the floor represent the ellipses \eqref{k.perp.s} for different values of $s$.
}
\end{figure}

Going back to Fig.~3 we see that the distribution of emitted photons 
is essentially supported on a straight line, $k'_\perp/k'_z = const$ (with an angular spread of the order $1/a_0 \ll 1$),
which intersects the ellipse (\ref{k.perp.s}) in a single point.
To relate back to the topic of depletion we recall Fig.~1 (right), which tells us that we have to stay away from the axes and the origin in the $a_0$-$\gamma_e$ plane according to our assumption of incoherent emission. The `safe' regime is thus $a_0 \sim \gamma_e \gg 1$, so that in terms of the emission angle we need to stay away from collinear emission, $a_0 \ll \gamma_e$ or $a_0 \gg \gamma_e$. Thus, in the generic regime of interest, $a_0 \sim \gamma_e \gg 1$, there is substantial transverse emission, cf.\ Fig.~3, right, for which the emission angle is about 50$^\circ$, with a depletion threshold of $N_T \approx 10^{10}$ according to Fig.~1.

In this letter we have reconsidered the multi-photon Compton process in strong EM fields, focussing on the energy loss of the laser due to absorption, which transforms the initially strong fields into weak ones. We found that this phenomenon has an intensity threshold of $a_0 \sim 10^3$, and requires $N_T \gtrsim 6.8\times 10^{11} \gamma_e^{0.92} a_0^{-1.08}$ electrons per laser wavelength cubed, according to the numerical fit in Fig.~2. We have neglected coherent photon emission, which is valid when $a_0 \sim \gamma_e \gg 1$.
It is expected that the depletion threshold will be overcome in the case of EM avalanches. Thus, laser depletion will not just be due to pair creation as considered previously, but must also be taken into account in laser photon absorption. 

We have further analyzed the photon emission rates differential in multi-photon number $s$ and discovered that they strongly peak at a value $s_0$, recall \eqref{s},  which determines the direction of the photon emission relative to the initial electron momentum direction in terms of an emission angle, $\theta$, via \eqref{tan-theta}. For generic depletion parameters, $a_0 \sim \gamma_e \gg 1$, one finds substantial emission in transverse direction.  In the collinear regime, $\theta \ll 1$ (forward scattering, $a_0 \ll \gamma_e$) and $\theta \approx \pi$ (back scattering, $a_0 \gg \gamma_e$), coherent emission can no longer be neglected.  Back scattering should dominate in the  EM avalanche regime, i.e.\ in colliding laser pulses or during interactions of laser pulses with solid density foils or plasmas of near-critical density. The classical interpretation of the emission angle $\theta$ in terms of averages over trajectories should yield a new test of the PIC codes currently in use.  

{In future work, we want to understand the effect of depletion on the emission probabilities. This will require estimating the effect of a decreasing $a_0$ on e.g.\ (\ref{eq:dPe_ds}), building on previous work such as \cite{Bergou:1981}.      
} 

\begin{acknowledgments}
We acknowledge support from the Office of Science of the US DOE under Contract No. DE-AC02-05CH11231. MM was supported by the Swedish Research Council grants \# 2012-5644 and 2013-4248. We would like to thank Anton Ilderton for fruitful discussions. The authors acknowledge the hospitality of the Kavli Institute for Theoretical Physics (KITP), where this research was initiated during the Frontiers of Intense Laser Physics Program and so was supported in part by the National Science Foundation under Grant No. NSF PHY11-25915.
\end{acknowledgments}

\end{document}